\documentclass[pre,twocolumn,superscriptaddress,floatfix,notitlepage,longbibliography]{revtex4-1}
\pdfoutput=1

\usepackage[pdftex]{color,graphicx,hyperref}
\usepackage{amsmath,amsfonts,amssymb}
\usepackage{bbding}
\usepackage{wasysym}

\usepackage[table,dvipsnames]{xcolor} 
\usepackage{colortbl}

\bibstyle{apsrev.bib}

\definecolor{plotcolor1}{HTML}{7AC36A}
\definecolor{plotcolor2}{HTML}{5A9BD4}
\definecolor{plotcolor3}{HTML}{FAA75B}
\definecolor{plotcolor4}{HTML}{9E67AB}
\definecolor{plotcolor5}{HTML}{737373}

\newcommand{\conv}{\ast}

\begin{document}

\title{\bf Estimating inter-event time distributions from finite observation periods in communication networks}
\author{Mikko Kivel\"a} 
\affiliation{Oxford Centre for Industrial and Applied Mathematics, Mathematical Institute, University of Oxford, Oxford OX2 6GG, UK}
\author{Mason A. Porter}
\affiliation{Oxford Centre for Industrial and Applied Mathematics, Mathematical Institute, University of Oxford, Oxford OX2 6GG, UK} 
\affiliation{CABDyN Complexity Centre, University of Oxford, Oxford OX1 1HP, UK}

\date{\today}

\begin{abstract}

A diverse variety of processes --- including recurrent disease episodes, neuron firing, and communication patterns among humans --- can be described using inter-event time (IET) distributions.  Many such processes are ongoing, although event sequences are only available during a finite observation window.  Because the observation time window is more likely to begin or end during long IETs than during short ones, the analysis of such data is susceptible to a bias induced by the finite observation period. In this paper, we illustrate how this length bias is born and how it can be corrected without assuming any particular shape for the IET distribution.  To do this, we model event sequences using stationary renewal processes, and we formulate simple heuristics for determining the severity of the bias.  To illustrate our results, we focus on the example of empirical communication networks, which are temporal networks that are constructed from communication events. The IET distributions of such systems guide efforts to build models of human behavior, and the variance of IETs is very important for estimating the spreading rate of information in networks of temporal interactions. We analyze several well-known data sets from the literature, and we find that the resulting bias can lead to systematic underestimates of the variance in the IET distributions and that correcting for the bias can lead to \emph{qualitatively} different results for the tails of the IET distributions.

\end{abstract}

\maketitle

\section{Introduction}

The newfound wealth of large data sets in the modern era of ``Big Data'' necessitates statistical analyses of such data. This has been prevalent in the study of human behavior, as the digital footprints left behind by electronic activities provide a deluge of data. One of the most important problems in the study of human dynamics, which benefits directly from such data, is to quantify temporal activity patterns in human behavior. For example, this problem has been approached via the characterization of time sequences of human activities~\cite{saramaki2015,Eckmann2004Entropy,Barabasi2005Origin,Vazquez2006Modeling,Rybski2012Communication,Candia2008Uncovering,Malmgren2008Poissonian,altmann2009,Wu2010Evidence,Karsai2011Small,Jiang2013Calling,Goh2008Burstiness,Kivela2012Multiscale,Duarte2007Traffic,Radicchi2009Human,jo2015} and the analysis of ``temporal networks''~\cite{Holme2012Temporal,holme13} (i.e., networks that change in time). \emph{Inter-event times} (IETs) give the times between each pair of events (e.g., sending an e-mail, making a phone call, or doing any other activity), and the way that they are distributed has received intense scrutiny because they can be used to characterize 
temporal processes.

Electronic records often have a huge number of data points. Such data often includes many subjects, but it may or may not also include a similar wealth of longitudinal points. For example, there exist data sets with thousands or even millions of people but with observation periods that only last a few months~\cite{Eckmann2004Entropy,Candia2008Uncovering,Karsai2011Small,Jiang2013Calling}.  
Moreover, even when the observation period is long, a given individual might rarely be active during that time.  This is the case, for example, in recent studies of e-mail communication~\cite{Eckmann2004Entropy,Vazquez2006Modeling,Malmgren2008Poissonian,Goh2008Burstiness,Barabasi2005Origin}, mobile phone calling~\cite{Karsai2011Small,Kivela2012Multiscale,Jiang2013Calling,saramaki2015}, website usage~\cite{Duarte2007Traffic,Radicchi2009Human,Rybski2012Communication}, and donations to charities \cite{wipprecht-thesis}.
As we will illustrate in this article, data sets in which the observation windows are comparable in scale to the IETs are vulnerable to finite-size biases. 
This can arise due to short observation windows and/or sparse records of activity.
This effect biases the tails of observed IET distributions, thereby creating a very serious issue, as the properties of distribution tails are often among the most important empirical features that one needs to consider \cite{Holme2012Temporal} and models of human dynamics have been validated or refuted based on their predictions of the shape of IET distributions \cite{Barabasi2005Origin,Vazquez2006Modeling,Malmgren2008Poissonian,Min2009Waiting,Oliveira2009Impact,Wu2010Evidence,Jiang2013Calling}. Furthermore, the variance of IET distributions can have a large effect on dynamical processes that occur on a system~\cite{Vazquez2007Impact,Karsai2011Small,Miritello2011Dynamical,Takaguchi2011Voter,Hoffmann2012Generalized,Kivela2012Multiscale,Jo2014Analytically,porter2014}, and the IET-distribution variance has been used to classify the processes that produce these distributions \cite{Goh2008Burstiness,Zhou2008Role}.

Several approaches have been used to account for the bias introduced by a finite temporal-window size. In particular, it is common to disregard all of the boundary effects and use the observed IETs~\cite{Duarte2007Traffic,Rybski2012Communication,Candia2008Uncovering,Goh2008Burstiness,Karsai2011Universal,Jiang2013Calling}. Such biases are sometimes acknowledged: for example, the exponential tail of an IET distribution is sometimes construed as a finite-size effect~\cite{Wu2010Evidence,Jiang2013Calling}. 
One can try to ameliorate the bias by introducing temporal periodic boundary conditions~\cite{Karsai2011Small,Kivela2012Multiscale}, but such a solution does not give an unbiased estimator for an IET distribution. Another approach to dealing with a finite observation period is to correct the probability of observing an IET value by dividing it by the probability that an IET of that length is not truncated by the observation window~\cite{Holme2003Network}. As we discuss in Section \ref{sec-estimate}, for stationary renewal processes, the latter probability always decreases linearly with the growth of the observed IET length. 
This linear correction has also been observed in empirical data by resampling using different observation window sizes~\cite{Vazquez2007Impact}.

The error in an observed IET distribution is very small if the tail of the IET distribution is sufficiently short relative to the length of the time window. This is usually the case if one uses event sequences with a large number of events. Unfortunately, in practice, this tends to entail that one can use only a small subset of available data. For example, some studies on temporal communication patterns that were based on data sets of thousands or millions of people only used subsets of the most active people that ranged from a single person to about $10\%$ of the data~\cite{Barabasi2005Origin,Wu2010Evidence,Malmgren2008Poissonian,Jiang2013Calling}. 
This approach discards valuable data and biases the analysis towards the behavior of very active individuals.

The use of IET distributions by scholars has a long history, and the problem of inferring an IET distribution from a finite observation period arises in a diverse set of fields --- such as engineering and medicine, where the problem has been studied using renewal processes~\cite{Feller1971Introduction} and other models for recurrent events~\cite{Nelson2003Recurrent,Cook2007Statistical}. 
Due to the generic nature of the problem, several statistical tools have been developed for estimating IET distributions for renewal processes~\cite{Vardi1982Nonparametric,Denby1985Shortcut,Mcclean1995Nonparametric,Soon1996Nonparametric,Pena2001Nonparametric,Gill2010Product,Zhu2014Parametric}. Additionally, some techniques based on survival analysis and event-history analysis have been used to analyze temporal network data~\cite{Butts2008Relational,Dubois2013Hierarchical,Melo2015Universal}. Similar problems have also been encountered when analyzing geological data~\cite{Laslett1982Censoring,Pickering1995Sampling} and estimating inter-spike intervals of firing neurons~\cite{Pawlas2011Distribution}.

\begin{figure}
\includegraphics[width=1.0\linewidth]{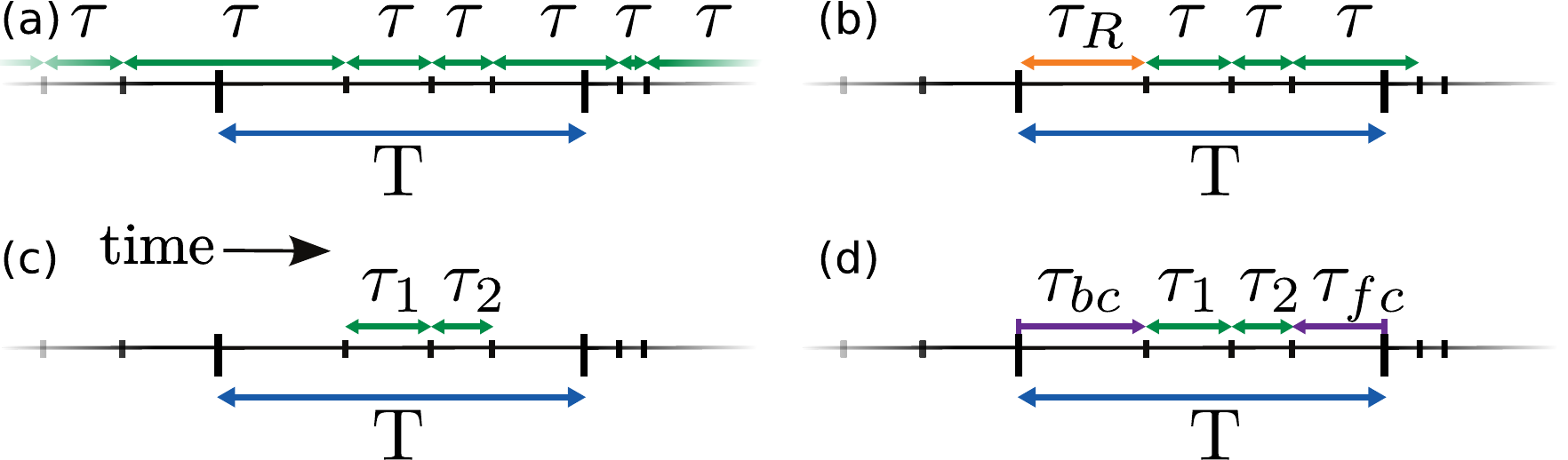}
\caption{(a) A stationary renewal process generates an infinite sequence of events. We place a time window of length $T$ in an arbitrary place on the timeline. (b) We consider only the events that lie inside of the time window. The time from the beginning of the time window to the first event is the \emph{residual waiting time} $\tau_R$, and one can derive its distribution from $p(\tau)$~\protect\cite{Feller1971Introduction}. (c) The \emph{observed inter-event times} are the IETs that lie completely inside of the time window. (d) The \emph{censored IETs} are the ones that are cut by the time window. 
An IET that is cut by the end (respectively, beginning) of the time window is said to be \emph{forward censored} (respectively, \emph{backward censored}). An IET that is truncated must be longer than the forward (and backward) censoring time: $\tau_{fc} \leq \tau_3$ (and $\tau_{bc} \leq \tau_0$) \protect\cite{modelfootnote}.
}
\label{fig:model_illustration}
\end{figure}

In the present paper, we concentrate on stationary renewal processes that produce $N$ event sequences observed in a finite time window of length $T$. See Fig.~\ref{fig:model_illustration}a,b for an illustration. We focus on renewal processes because they are minimal models for producing event sequences with arbitrary IET distributions. However, real processes are often more complicated than stationary renewal processes. For example, communication patterns and many natural phenomena --- such as earthquakes, neuronal spike trains, and disease epidemics---arise from processes that have memory~\cite{Goh2008Burstiness,Karsai2011Universal}.
Other processes, such as inhomogeneous Poisson processes and processes in which cascades of activity can be triggered by prior events, also yield tractable models for human dynamics~\cite{Malmgren2008Poissonian,Malmgren2009Universality,Zipkin2014Pointprocess}.

\section{Estimation of inter-event time distributions} \label{sec-estimate}

We seek to estimate the IET distribution $p(\tau)$ of the underlying process when we are given only the time stamps of the events inside of the observation window. A naive method would be to use the distribution $p^\prime(\tau)$ for observed IETs to estimate the real distribution $p(\tau)$ (see Fig.~\ref{fig:model_illustration}c). Unfortunately, in general, the observed IETs and the real IETs do not follow the same distribution.  

\begin{figure}
\includegraphics[width=1.0\linewidth]{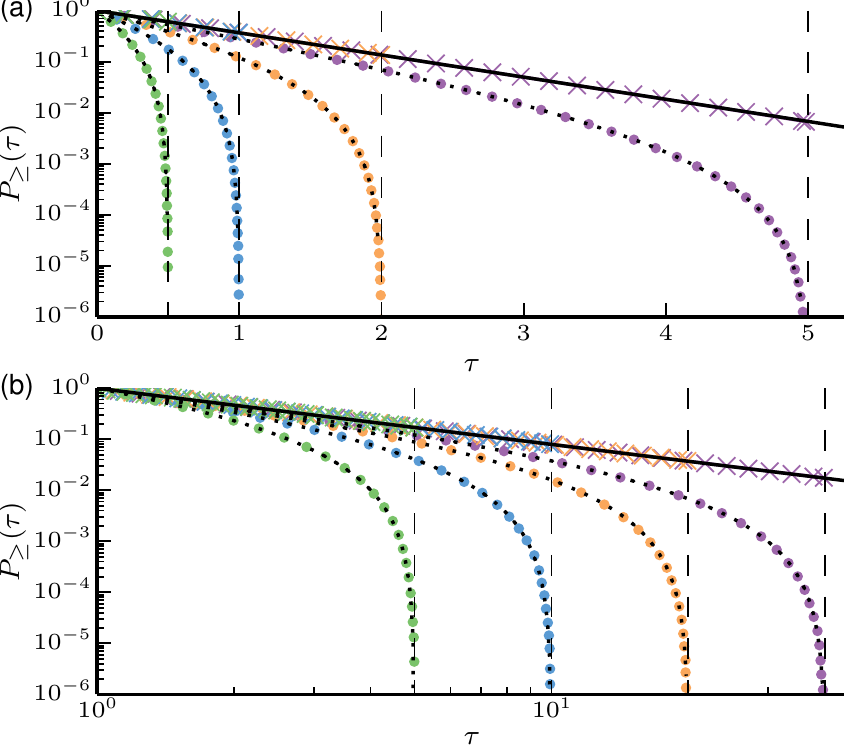}
\caption{
We simulate $N=10^5$ event sequences using stationary renewal processes for which the real IET distribution satisfies (a) $p(\tau) \propto e^{-\tau}$ and (b) $p(\tau) \propto \tau^{-2.1}$. We consider window sizes $T$ (which we indicate with dashed vertical lines) of (a) 0.5, 1, 2, and 5 and (b) 5, 10, 20, and 40. 
We calculate IETs for each event sequence, pool them together, and plot cumulative IET distributions $P_{\ge}(\tau)$. The dots indicate the observed IET distribution, and the crosses indicate the estimates of the real IET distribution using the Kaplan--Meier (KM) estimator. The solid black line is the theoretical $p(\tau)$ distribution, and the dotted curves are the theoretical distributions $p^\prime(\tau)$ for IETs [see Eq.~\protect\eqref{eq:tauobs}]
that lie completely inside of each time window.
A nonparametric maximum likelihood estimator (NPMLE)~\protect\cite{Soon1996Nonparametric} gives qualitatively similar results. See Fig.~\protect\ref{fig:model_iets_binned} in Appendix \protect\ref{sec:altplots} for the same distributions plotted using probability densities instead of cumulative probabilities.
}
\label{fig:model_iets}
\end{figure}

In Fig.~\ref{fig:model_iets}, we illustrate the difference between $p(\tau)$ and $p^\prime(\tau)$ for stationary renewal processes with exponential and power-law IET distributions. This difference grows linearly when the IET length $\tau$ approaches the window size $T$, and $p^\prime(\tau)=0$ for $\tau > T$. The growth occurs because a longer IET makes it more likely that the observation window either starts or end between the two events that correspond to that IET. Observed IETs are always distributed so that there is a linear cutoff at the end time $T$ of the time window. In other words, 
\begin{equation}
	p^\prime(\tau) \propto (T - \tau) p(\tau)
\label{eq:tauobs}
\end{equation}
when the number $N$ of event sequences tends to infinity \cite{Soon1996Nonparametric}.
To give intuition for Eq.~\eqref{eq:tauobs}, note for a stationary renewal process that the probability of observing an event is uniform for a whole observation window. This implies that $(T - \tau)/T$ is the probability that an IET of length $\tau$ following an event chosen uniformly at random in the interval is not cut short by the end of the observation window.

In the worst case, the linear bias in Eq.~\eqref{eq:tauobs} can lead to qualitatively incorrect conclusions about the shape of the tail of an IET distribution. It is therefore important to correct for this bias.
Note that this bias is more severe than that from an upper truncation, in which data points that are larger than a certain threshold value are not
observed~\cite{Burroughs2001Upper,Deluca2013Fitting}.

There exist both parametric~\cite{Zhu2014Parametric} and nonparametric~\cite{Vardi1982Nonparametric,Denby1985Shortcut,Mcclean1995Nonparametric,Soon1996Nonparametric,Pena2001Nonparametric,Gill2010Product,Cook2007Statistical} estimators for the real IET distribution $p(\tau)$. A straightforward nonparametric way to estimate IETs is to use the Kaplan--Meier (KM) estimator~\cite{Kaplan1958Nonparametric} by considering the IETs inside of the time window as uncensored observations and the IETs that are truncated by the end of the time window as censored observations~\cite{Gill1981Testing,Denby1985Shortcut}. Additionally, because the stationary renewal process that generates the event sequences is symmetric in time, we can increase the accuracy of our estimate by repeating this estimation process backwards in time~\cite{Denby1985Shortcut}. That is, each uncensored IET is counted twice, and the censored IETs at the boundaries of the time window are counted only once \footnote{We call this estimator ``the KM estimator'' in the rest of our article. It is also sometimes called a ``a product-limit estimator''}. One can estimate the variance of the KM estimator using Greenwood's formula~\cite{Kaplan1958Nonparametric,Greenwood1926Natural}, which has to be modified slightly to take into account double-counting of the uncensored IETs~\cite{Denby1985Shortcut}. See Fig.~\ref{fig:model_illustration}d for a schematic and Fig.~\ref{fig:model_iets} for an example how the KM estimator corrects the bias introduced by the finite observation window for simulated data \cite{ietsoft}. See Appendix \ref{sec:km_estimator} for details on how to use the KM estimator to estimate IET distributions.

The derivation of the KM estimator for IETs is based on a partial likelihood approach for data produced with a stationary renewal process~\cite{Denby1985Shortcut}. The KM estimator only assumes that the sampled IETs are produced from the IET distribution independently of the windows of observation (i.e., the times from events to the end of the observation period). That is, the KM estimator disregards some information on how the data were produced if it is used for data that is known to be produced by a stationary renewal process.
Vardi~\cite{Vardi1982Nonparametric} defined a nonparametric maximum likelihood estimator (NPMLE) method for data produced with a stationary renewal process. Soon and Woodroofe~\cite{Soon1996Nonparametric} later generalized Vardi's method for continuous-time situations as well as for situations in which there are event sequences in which no events are observed during the observation time window. Note, however, that methods based on the KM estimator and Vardi's NPMLE can yield estimates that are very close to each other even though the KM estimator is more computationally efficient than Vardi's NPMLE estimator~\cite{Denby1985Shortcut}. One can also use a reduced-sample estimator, which ignores data points close to the boundaries of an observation window, although Pawlas et al. \cite{Pawlas2011Distribution} observed for several different generative models of event sequences that it gives less accurate estimates than a method based on the KM estimator.

\subsection{When does one need to worry about finite window-size effects?} \label{subsec-worry}

The bias introduced by using the observed IET distribution as an estimate for the real IET distribution for a given process can be very small even if data are produced by sampling from a renewal process using a finite time window. This is the case if the time-window length is sufficiently long. 
In this case, one does not need to worry about finite-size effects or make any corrections to account for them. We will next give some guidelines for determining when this happy situation holds.

As we discussed at the beginning of Section~\ref{sec-estimate}, the bias in an IET distribution grows linearly with IET length. It is thus useful to compare the bias in the smallest observed IET to the bias in the largest observed IET, as their ratio gives an estimate for the largest error in the distribution. If the smallest possible IET is $\tau_0$, then Eq.~\eqref{eq:tauobs} implies that
\begin{equation}
	\frac{p^{\prime}(\tau )}{p^{\prime}(\tau_0)} =\left(1-\frac{\tau}{T}\right) \frac{p(\tau )}{p(\tau_0)}\,.
\label{eq:rot}
\end{equation}
Equation \eqref{eq:rot} can be used as a rule of thumb for assessing if a finite time window distorts an observed IET distribution.
For example, if the largest data point (i.e., the rightmost point in an observed IET distribution) is more than $100$ times smaller than the length of the observation window, then the error that results using the observed IETs for estimating the real IET distribution is less than $1\%$ for IET values that are smaller than the maximum observed IET value.

Equation (\ref{eq:rot}) gives an estimate for the relative probabilities of observed IETs, but it does not indicate anything about the distribution's tail, which is not observed. This can be an issue if there are very small amounts of data or if one wants to calculate summary statistics of an IET distribution that are very sensitive to the properties of the tail (e.g., moments of an IET distribution, measures of event burstiness \cite{Goh2008Burstiness}, and so on).  The moments $\mu_m^\prime$ of an observed IET distribution are lower than the moments $\mu_m$ of the real IET distribution. However, if we have an estimate $p^{\rm{est}}( \tau )$ for the real IET distribution $p( \tau )$ for $\tau \leq \tau_{\rm{max}}$, then we can obtain an estimate for the moments using
\begin{align}
	\mu_{m}^{\rm{est}}=\int_0^{\tau_{\rm{max}}} \tau^m p^{\rm{est}}( \tau )d\tau + \tau_{\rm{max}}^m  P_{\ge}^{\rm{est}}(\tau_{\rm{max}})\,,
\label{eq:muest}
\end{align}
where $P_{\ge}^{\rm{est}}$ is the estimator of cumulative distribution of the IETs. 
That is, in this estimator, we use $p^{\rm{est}}( \tau )$ for the IET distribution for $\tau \leq \tau_{\rm{max}}$ and replace the unobserved tail by adding all of the remaining probability mass, $P_{\ge}^{\rm{est}}(\tau_{\rm{max}})$, to the point $\tau_{\rm{max}}$. Assuming that the estimate for the IET distribution is perfect (i.e., $p^{\rm{est}}( \tau )=p( \tau )$ when $\tau \leq T$), we obtain a sharper lower bound for the moments using $\mu_{m}^{\rm{est}}$ than using $\mu_m^\prime$. That is,  $\mu_m^\prime \leq \mu_{m}^{\rm{est}} \leq \mu_{m}$. We illustrate this issue in Section~\ref{sec:data} using empirical data.  Note, in practice, that $\tau_{\rm{max}}$ is close to $T$ for these data sets.

\section{Analysis of empirical data} 
\label{sec:data}

We now use the methods that we described in Section \ref{sec-estimate} to reanalyze several public data sets that have been studied previously in the literature. For each data set, we concentrate on temporal sequences of messages that are sent by individuals. 

The Eckmann et al. e-mail data set~\cite{Eckmann2004Entropy} contains time stamps of about $3 \times 10^5$ e-mails between 3188 people during $83$ days. This data set has been examined by several authors, and the shape of the IET distributions of individuals with high e-mailing frequencies has received particularly close scrutiny (and has attracted controversy)~\cite{Barabasi2005Origin,Stouffer2005Comment,Barabasi2005Reply,Malmgren2008Poissonian,Karsai2011Universal}.
The \url{pussokram.com} (POK) data set~\cite{Holme2003Network,Rybski2012Communication} is a communication record of an online community with about $3 \times 10^4$ people who sent $5 \times 10^5$ messages during the entire 492-day lifetime of the site. Because the data recording started from the birth of the POK website, it is not reasonable to construe message sequences in this data set as having been produced by a stationary process. However, it is still reasonable to consider the data as being forward censored (see Fig.~\ref{fig:model_illustration}). Rybski et al.~\cite{Rybski2012Communication} plotted the distribution of all IETs as well as distributions grouped according to the number of sent messages.  Their plots contain noticeable dips at the end of the IET distributions, but it is not clear in their paper if this feature arises because of intrinsic human behavior or is instead due to the finite length of the data. A third data set that we examine was introduced by Wu et al.~\cite{Wu2010Evidence}, who studied IETs of short messages sent within three different companies during one month. We present our reanalysis of data from company 1, which includes about $5 \times 10^5$ messages sent by about $4 \times 10^4$ people. The results for the two other companies are similar.
To obtain good statistics, Wu et al. concentrated on communication patterns between the few pairs of users who sent very large numbers of messages to each other.  
For each data set, we consider the observation window for each user to be the observation window of the whole system, although additional information of users leaving or joining the system could have been used to construct individual observation windows if such information were available.

\begin{figure}
\includegraphics[width=1.0\linewidth]{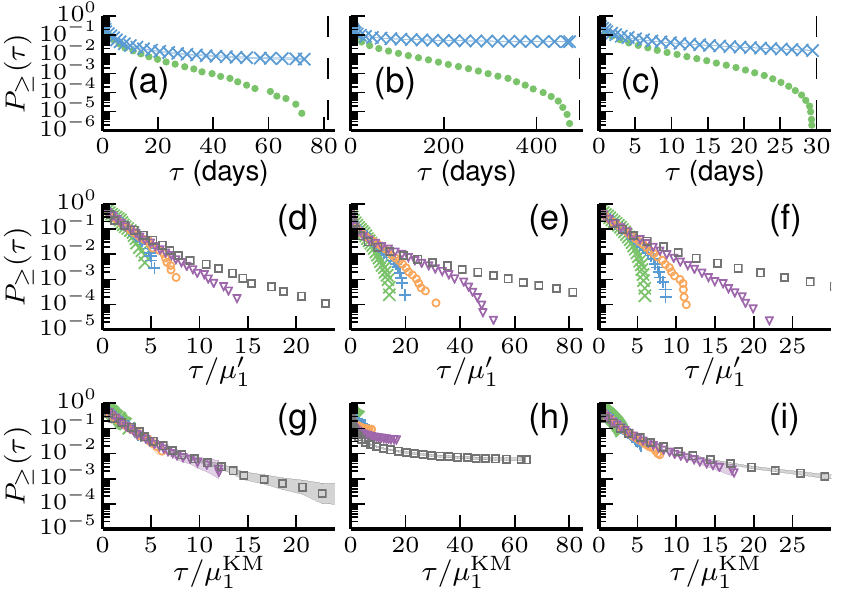}
\caption{Results for the empirical data sets. We consider activation times for each node as a single event sequence.
In panels (a)--(c), we show the IET distributions that we obtain from combining IET distributions of all node activation sequences. The dots indicate the observed IET distributions, and the crosses indicate the estimates of the IET distributions using the KM estimator.
 In panels (d)--(i), we bin the event sequences according to the number of events in them ({\protect\color{plotcolor1}$\times$}: $n = 3$, {\protect\color{plotcolor2}$+$}: $n = 6$, {\protect\color{plotcolor3}$\circ$}: $n \in [8, 9]$, {\protect\color{plotcolor4}$\bigtriangledown$}: $n \in [14, 25]$, {\protect\color{plotcolor5}$\square$}: $n \in [51, 150]$). We skip every other bin to make the figure easier to read), and we normalize 
 the IETs
 according to the bin's mean IET.
In panels (d)--(f), we show cumulative distributions of observed IETs normalized by the mean $\mu_1^\prime$ of the observed IETs. In panels (g)--(i), we show KM estimates for the cumulative IET distributions normalized by the mean $\mu_1^{\rm{KM}}$ calculated from the estimated IET distribution.
The shaded regions are the 95\% confidence intervals~\protect\cite{Denby1985Shortcut}.
The data sets are (a, d, g) the Eckmann et al. e-mail data~\protect\cite{Eckmann2004Entropy}, (b, e, h) POK messages~\protect\cite{Rybski2012Communication}, and (c, f, i) the Wu et al. short-message data~\protect\cite{Wu2010Evidence}.
See Fig. \protect\ref{fig:data_iets_loglog} in Appendix \protect\ref{sec:altplots} for the same distributions plotted using doubly logarithmic axes.
} 
\label{fig:data_iets}
\end{figure}

Each of the data sets includes a large number of IETs that are sufficiently close to the time-window length to affect the observed IET distribution.  We illustrate this fact in panels (a)--(c) of Fig.~\ref{fig:data_iets}.  For each data set, we show both the observed IET distribution and the KM estimate of the IET distribution. It is clear that the shape of the tail of the observed IET distributions is qualitatively different from that of the KM estimate of the IET distribution. The dip that is often observed in the tail of an IET distribution that includes IETs that are close to the observation-window length~\cite{Duarte2007Traffic,Karsai2011Universal,Rybski2012Communication,Jiang2013Calling} can be explained by the finite observation window in each of the data sets that we study.

In Table~\ref{table:summary}, we compare some summary statistics of the KM estimate of the IET distributions and the observed IET distributions to gain a better understanding of how much the two differ.
The first two moments and residual waiting times calculated from the IET distribution given by the KM estimator are often more than $100\%$ larger than ones calculated from the observe IET distribution. 
These differences can have a huge impact on processes that act on top of temporal networks, and it is clear that the bias introduced by a finite observation-window size can be a major problem in these situations.
For example, the mean residual waiting time $\tau_R$ --- which is vastly smaller when calculated using the observed IET distributions than when calculated using the IET distributions obtained with the KM estimator --- is related to the speed of spreading in networks~\cite{Vazquez2007Impact,Karsai2011Small,Miritello2011Dynamical,Kivela2012Multiscale,Jo2014Analytically}, because it is the expected time 
until the next event after a node is infected at a time chosen uniformly at random.

\begin{table}
\begin{tabular}{l|cc|cc|ccc}
Data&$\mu_1^\prime$&$\mu_1^{\rm{KM}}$&$\sqrt{\mu_2^\prime}$&$\sqrt{\mu_2^{\rm{KM}}}$&$\mu_1^\prime (\tau_R)$&$\mu_1^{\rm{KM}} (\tau_R)$&$\tau_{fc,bc}$\\
\hline
E-mail & $0.908$&$1.51$&$3.20$&$6.88$&$5.62$&$15.6$&$17.5$ \\
POK & $5.13$&$28.4$&$23.1$&$106$&$51.9$&$198$&$240$ \\
Short message & $0.633$&$1.40$&$2.11$&$4.89$&$3.53$&$8.53$&$8.73$ \\
\end{tabular}
\caption{The first two moments of the IETs calculated from the observed IET distribution and using Eq.~\protect\eqref{eq:muest} for the IET distribution produced by the KM estimator, estimates of the residual waiting times using the formula $\mu_1(\tau_R)=\frac{1}{2}\frac{\mu_2}{\mu_1}$~\protect\cite{Kivela2012Multiscale}, and the mean of forward and backward censoring times $\tau_{fc,bc}$. (For the POK data set we only calculate the mean of the forward censoring times.) 
Note that data produced by a stationary renewal process has forward-censoring and backward-censoring times that are are distributed as the residual waiting times for values that are smaller than the window size $T$.
}
\label{table:summary}
\end{table}

In studies of empirical data, it is often assumed that each event sequence is produced by an IET distribution with the same characteristic shape but different underlying rate. Different event sequences would then arise using the same scaling function $f$ but with a different mean value $\tau_0$ \cite{Corral2003Local,Saichev2006Universal,Candia2008Uncovering,Goh2008Burstiness,Karsai2011Small,Rybski2012Communication}. The IET distribution for a sequence with finite mean $\tau_0$ is defined as $p(\tau | \tau_0) = \frac{1}{\tau_0} f(\tau / \tau_0)$ (see Appendix \ref{sec:scaling}). In panels (d)--(i) of Fig.~\ref{fig:data_iets}, we plot the IET distributions (for each data set) in which we group event sequences with similar numbers of events. We include event sequences that have fewer than 151 events because sequences with few events are the most susceptible to finite-size effects. Sequences with at most 150 events encompass 90\%--99\% of all sequences (depending on the data set). We observe that normalized IET distributions for event sequences with few events decrease much faster than the IET distributions for sequences with many events. This result is expected, and it results from the bias introduced by the finite observation window. There is a very good collapse of the tails of the KM estimates of the normalized IET distributions for the e-mail communication and short-message communication data. This is remarkable, given that collapse is not expected to be perfect even for data that perfectly follows the characteristic distribution model (see Appendix \ref{sec:scaling}). The difference between the IET distributions of the POK data and the two other data sets may be due to users who leave the service permanently. This process would lead to the last IET being infinitely long, which would manifest as the tail of the cumulative distribution approaching a value that corresponds to the fraction of people in each group who have left the service. 
One would expect this fraction to be smaller for groups with a larger number of messages if the probability of leaving the service is lower for people who have sent more messages.

\section{Conclusions and Discussion} 

We investigated the effects that a finite observation window can have on observed inter-event times (IETs). For a stationary renewal process, we illustrated that the finite time window introduces a linear cutoff to the observed IET distribution at the end of the time window (see Fig.~\ref{fig:model_iets}). We showed  how to correct this bias using nonparametric estimators, such as the KM estimator or an NPMLE, for a stationary renewal process. We also illustrated that these estimators work well even for event sequences with small numbers of events if these sequences can be grouped together.
We then used these methods to reanalyze three data sets of human communication, and we found that using the observed IET distributions without correcting for the finite-size bias can seriously distort the shape and key summary statistics of IET distributions.

Human behavior is rather heterogeneous in many aspects, and in particular, the event sequences of different people contain widely disparate number of events. Many authors have argued that it is possible to represent such sequences using a scaling function that is independent of the underlying rate of events~\cite{Corral2003Local,Saichev2006Universal,Candia2008Uncovering,Goh2008Burstiness,Karsai2011Small}. However, there is an additional bias if one infers the underlying rate from the observed number of events (see Appendix~\ref{sec:scaling}), and it is important to develop statistical methods that are able to assume an underlying model for a characteristic IET distribution. Moreover, methods for testing whether an IET distribution has some specific shape are also susceptible to finite-size effects, and parametric analogs of the methods that we have employed should be applied in such situations~\cite{Zhu2014Parametric}. Further, in the present paper we are focusing on the IET distributions of multiple event sequences, but finite-size effects should also be taken into consideration when estimating summary statistics such as moments or burstiness \footnote{Note that the burstiness coefficient defined in Ref.~\cite{Goh2008Burstiness} is based on the coefficient of variation, and naive estimates of the coefficient of variation are biased for small sample sizes~\cite{Sokal1980Significance,Breunig2001Almost}.} of single event sequences.

The need for the wide dissemination and use of correction methods like KM estimators or NPMLEs for IET distributions is underscored by the rapidly growing analysis of temporal data streams. Nonparametric methods for correcting for biases that are introduced by a finite observation window have existed for several decades~\cite{Vardi1982Nonparametric,Denby1985Shortcut,Soon1996Nonparametric}. Surprisingly, such methods (to our knowledge) do not seem to have been used when analyzing human communication patterns, although there have been some ad-hoc attempts to directly correct for the linear bias~\cite{Holme2003Network,Vazquez2007Impact}. Additionally, although we have focused on human communication patterns, the problem of correcting for these finite-size effects is a general one, and similar methods have been reinvented in multiple fields. For example, the KM estimator was used for window-censored data in the 1980s~\cite{Denby1985Shortcut}, and its use for such data was independently reinvented many years later in the context of estimating the inter-spike intervals of neurons~\cite{Pawlas2011Distribution}. Appropriately taking into account finite-size effects makes it possible to obtain accurate estimates for the tail of an IET distribution and to optimally exploit data that consists of a large number of event sequences with only a small number of events (as opposed to high-frequency event sequences, which are largely free of such significant finite-size effects).

\begin{acknowledgments}
Both authors were supported by the European Commission FET-Proactive project PLEXMATH (Grant No. 317614).  We thank Andrea Bertozzi, Carlos Gershenson, Adilson Motter, Se Wook Oh, and Jari Saram\"aki for helpful comments; and we thank Jean-Pierre Eckmann for providing us with the e-mail data set. We also thank several anonymous referees for helpful comments.
\end{acknowledgments}

\appendix
\section{Kaplan--Meier estimator for inter-event times}\label{sec:km_estimator}

We now discuss how to use the Kaplan--Meier (KM) estimator~\cite{Kaplan1958Nonparametric} to estimate the IET distribution of a stationary renewal process when one only observes events in a finite time window. Our approach is similar to the ``shortcut method'' of Denby and Vardi~\cite{Denby1985Shortcut}.  Unlike them, however, we do not add a point $\tau_M$ that is much larger than the observed IET values to the IET-distribution estimate.

The KM estimator is a nonparametric estimator for lifetimes (or times of death) in the presence of censored lifetimes (or losses)~\cite{Kaplan1958Nonparametric}. Corresponding to each lifetime $\tau_i$, there is a censoring time $\tau_{c,i}$, and we observe the lifetime if it is shorter than or the same as the censoring time (i.e., if $\tau_i \leq \tau_{c,i}$) and censor it if it is longer than the censoring time (i.e., if $\tau_i > \tau_{c,i}$). That is, for each $i$ we observe a single time $t_i$ that is either a lifetime $t_i=\tau_i$ or a censoring time $t_i=\tau_{c,i}$. (If the lifetime $\tau_i$ is censored, we say that it is a ``censored lifetime,'' and we say that the time $\tau_{c,i}$ that it is censored is its ``censoring time.'')  The KM estimator $\hat{P}_{\ge}$ for the cumulative distribution of lifetimes is
\begin{align}
	\hat{P}_{\ge}(t)= \prod_{s \le t }\left(1- \frac{\delta_s}{n_s}\right)\,,
\end{align}
where $n_s$ is the number of lifetimes that are known to be at least as long as $s$ (i.e., $n_s=\sum_{s^\prime \geq s}[\delta_{s^\prime}+c_{s^\prime}]$), the parameter $\delta_s$ is the number of lifetimes that are observed at time $s$, and $c_s$ is the number of lifetimes that are censored at time $s$.

One can estimate the variance of the KM estimator using Greenwood's formula~\cite{Kaplan1958Nonparametric,Greenwood1926Natural}:
\begin{align}
	\mathrm{Var}\left(\hat{P}_{\ge}(t)\right)=\hat{P}_{\ge}^2(t) \sum_{s \le t }\frac{\delta_s}{n_s(n_s - \delta_s)}\,.
\label{eq:greenwood}
\end{align}
One can then use the variance estimate to construct confidence intervals for the estimate of an IET distribution. For example, if the $\hat{P}_{\ge}(t)$ values are normally distributed, then the confidence intervals are 
\begin{equation*}
	\hat{P}_{\ge}(t) \pm z_{\alpha/2} \sqrt{\mathrm{Var}\left(\hat{P}_{\ge}(t)\right)}\,, 
\end{equation*}	
where $1-\alpha$ is the confidence level and $z_{\alpha}$ is the quantile function of the standard normal distribution. 
In general, however, the $\hat{P}_{\ge}(t)$ values are not normally distributed, which can lead to confidence intervals that are not restricted to lie in the interval $[0,1]$. One usually addresses this situation by applying a transformation $g$ to the $\hat{P}_{\ge}(t)$ values to obtain a set of values that better follow a normal-distribution approximation. One can then calculate the confidence interval for the transformed random variable so that 
\begin{equation*}
	g\left(\hat{P}_{\ge}(t)\right) \pm z_{\alpha/2} \sqrt{\mathrm{Var}\left(g\left(\hat{P}_{\ge}(t)\right)\right)}\,. 
\end{equation*}	
Choices for the transformation include $g(p)=\ln(p)$, $g(p)=\ln\left(-\ln(p)\right)$, and $g(p)=\arcsin(\sqrt{p})$. (See, e.g., Borgan and Knut~\cite{Borgan1990Note} for a discussion about choosing the transformation.) In Fig.~\ref{fig:data_iets} of the main text, we used the transformation $g(p)=\ln\left(p/(1-p)\right)$ to follow the choice in Ref.~\cite{Denby1985Shortcut}. 

One can use the KM estimator to estimate IETs of a renewal process by considering the observed IETs as observed lifetimes and the IETs that are truncated by the end of a time window (i.e., the IETs that are forward censored) as censored lifetimes. If the renewal process is stationary, then one can also repeat this procedure by reversing the direction of time~\cite{Denby1985Shortcut}. In other words, one can consider both backward-censoring and forward-censoring times as censored lifetimes, and the observed IETs are twice counted as observed lifetimes. This makes it possible to use the information in the backward-censoring times in the construction of the estimator for the IET distributions. Note that the variance estimator of Greenwood's formula in Eq.~\eqref{eq:greenwood} needs to be multiplied by $2$ in order to account for the fact that uncensored data points are used twice~\cite{Denby1985Shortcut}.

\section{Alternative illustrations of IET distributions}\label{sec:altplots}

 \begin{figure}
\includegraphics[width=1.0\linewidth]{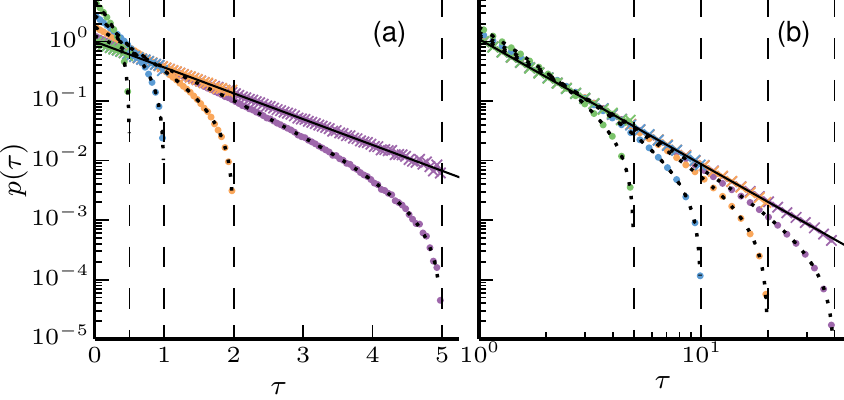}
\caption{As in Fig.~\protect\ref{fig:model_iets}, we simulate $N=10^5$ event sequences using stationary renewal processes. Now, however, we plot probability densities instead of cumulative probabilities. We plot IET distributions $p(\tau)$ for $N=10^6$ event sequences that we simulate from a stationary renewal process for which (a) $p(\tau) \propto e^{-\tau}$ and (b) $p(\tau) \propto \tau^{-2.1}$. We consider window sizes $T$ (which we indicate with dashed vertical lines) of (a) 0.5, 1, 2, and 5 and (b) 5, 10, 20, and 40. The dots indicate the observed IET distribution, and the crosses indicate the estimates of the real IET distribution using the KM estimator. The solid black line is the theoretical $p(\tau)$ distribution, and the dotted curves are the theoretical distributions $p^\prime(\tau)$ for IETs [see Eq.~\eqref{eq:tauobs}] that lie completely inside of each time window.
}
\label{fig:model_iets_binned}
\end{figure}

\begin{figure}
\includegraphics[width=1.0\linewidth]{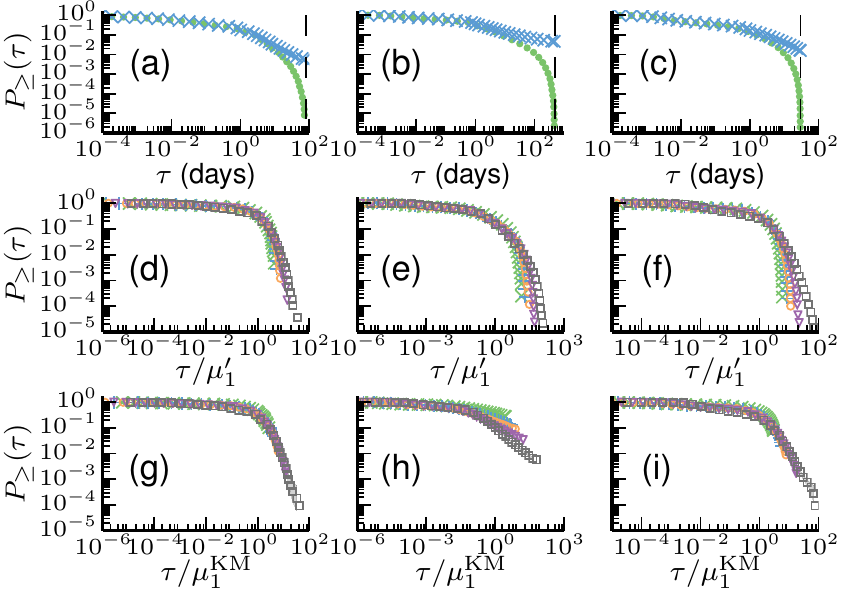}
\caption{Results for the empirical data sets (also see Fig.~\ref{fig:data_iets}) plotted using doubly logarithmic axes. 
 We consider activation times for each node as a single event sequence.
In panels (a)--(c), we show IET distributions that we obtain by combining IET distributions of all node activation sequences. 
The dots indicate the observed IET distributions, and the crosses indicate the estimates of the IET distributions using the KM estimator.
 In panels (d)--(i), we bin the event sequences according to the number of events in them ({\protect\color{plotcolor1}$\times$}: $n = 3$, {\protect\color{plotcolor2}$+$}: $n = 6$, {\protect\color{plotcolor3}$\circ$}: $n \in [8, 9]$, {\protect\color{plotcolor4}$\bigtriangledown$}: $n \in [14, 25]$, {\protect\color{plotcolor5}$\square$}: $n \in [51, 150]$). We skip every other bin to make the figure easier to read, and we normalize the IETs according to the bin's mean IET.
 In panels (d)--(f), we show cumulative distributions of observed IETs normalized by the mean $\mu_1^\prime$ of observed IETs. In panels (g)--(i), we show KM estimates for the cumulative IET distributions normalized by the mean $\mu_1^{\rm{KM}}$ calculated from the estimated IET distribution.
The shaded regions are the 95\% confidence intervals~\protect\cite{Denby1985Shortcut}.
The data sets are (a, d, g) the Eckmann et al. e-mail data~\protect\cite{Eckmann2004Entropy}, (b, e, h) POK messages~\protect\cite{Rybski2012Communication}, and (c, f, i) the Wu et al. short-message data~\protect\cite{Wu2010Evidence}.
} 
\label{fig:data_iets_loglog}
\end{figure}

Figure \ref{fig:model_iets_binned} corresponds to Fig.~\ref{fig:model_iets} in the main text, but we now show probability densities instead of cumulative probabilities. Figure \ref{fig:data_iets_loglog} corresponds to Fig.~\ref{fig:data_iets} in the main text, but we now plot the IET distributions using doubly logarithmic axes.

\section{Analyzing event sequences selected based on the number of events in them}
\label{sec:scaling}

\subsection{Distributions of number of events} \label{sec:nevents}

\begin{figure}
\includegraphics[width=0.49\linewidth]{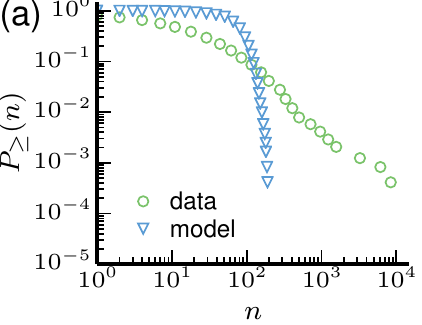}
\includegraphics[width=0.49\linewidth]{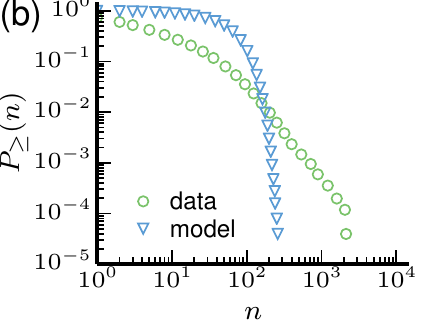}
\includegraphics[width=0.49\linewidth]{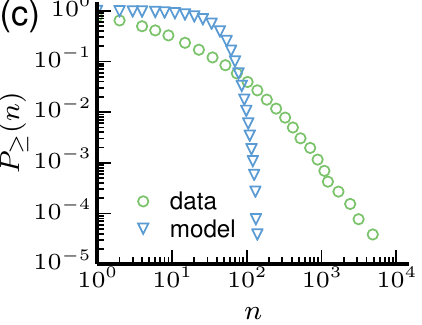}
\caption{Cumulative distributions for the numbers of events in several communication data sets. We indicate the distribution of the original data using green circles, and we use blue triangles to indicate the distribution of the process that assumes that the data were produced by a single IET distribution. (See the text for details.) (a) Eckmann et al. e-mail data~\protect\cite{Eckmann2004Entropy}, (b) POK messages~\protect\cite{Rybski2012Communication}, and (c) Wu et al. short-message data \protect\cite{Wu2010Evidence}.
} 
\label{fig:nevents}
\end{figure}

One can quantify the activity of the people in 
the data sets discussed in the main text
by counting the number of events that each person has in his/her event sequence.
Most of the people in the data that we examine exhibit very little activity, although there are also people that are significantly more active (by several orders of magnitude). One would not expect such a distribution if all event sequences were produced by a single renewal process. To illustrate this point, we construct a renewal process whose IET distribution we infer using the KM estimator. (See Fig.~\ref{fig:data_iets} in the main text.) Using this model process, we produce a new data set that has the same number of event sequences as the original data. In Fig.~\ref{fig:nevents}, we plot the activity distribution for the original data and the data produced by the model processes. The distributions of events observed in our data sets and the ones observed for the model are significantly different:
almost all of the event sequences produced by the renewal process that we construct contain between $10$ and $100$ events, and there are no sequences with a very small number or a very large number of events. It is clearly very unlikely that all of the event sequences in the data were produced by a single renewal process.

\subsection{Model with a scaling function}\label{app:universal}

One way to relax the assumption that event sequences are produced by a single IET distribution is to suppose that each event sequence is produced by an IET distribution with the same characteristic shape, which given by a scaling function $f$ but with a different mean value $\tau_0$. The IET distribution for a model constructed using this scenario is $p(\tau | \tau_0) = \frac{1}{\tau_0} f(\tau / \tau_0)$, where $\tau_0$ is the mean IET of the sequence. 
Such a model has been fitted to several empirical data sets \cite{Corral2003Local,Saichev2006Universal,Candia2008Uncovering,Goh2008Burstiness,Karsai2011Small,Rybski2012Communication}.

Let's consider a model in which we choose the distributions $f$ and $p_0(\tau_0)$ so that our model resembles a real set of event sequences but remains analytically tractable. The distribution for the number of events is often heavy-tailed in communication data~\cite{Onnela2007Analysis} (e.g., see Fig.~\ref{fig:nevents}), and we choose to model the distribution for the number of events as $p(n) \propto n^{-\alpha}$ (where $n \geq 1$ and $\alpha=2.5$). To do this, we construct the distribution $p_0$ for the mean values $\tau_0$ so that the numbers of events in the sequences are distributed as the given power law. To ensure analytical tractability, we choose the function $f$ to be an exponential function. That is, our aggregate process is a combination of multiple Poisson processes.

For each event sequence, we draw an expected IET from the distribution $p_0(\tau_0)$. Event sequences are then produced by a renewal process with an IET distribution of $p(\tau)=f(\tau / \tau_0)/\tau_0$. The residual waiting-time distribution~\cite{Feller1971Introduction} for the process is then
\begin{align}
	p_R(\tau_R) = \frac{1}{\tau_0} f_R(\tau_R/\tau_0)\,,
\label{puniversal}
\end{align}
where $f_R$ is the residual waiting-time distribution for the process that is determined by the IET distribution $f$. By exploiting the expected relation $n=\frac{T}{\tau_0}$, we can approximate the IET distribution for the aggregate process:
\begin{align}
	p(\tau ) \propto \int_1^{\infty} n p_0(n) p\left(\tau | \tau_0=\frac{T}{n}\right) dn\,,
\end{align}
which reduces to
\begin{align}
	p(\tau ) \propto E_{\alpha - 2}(\tau / T)\,,
\label{eq:iet_collapse_plaw_exp}
\end{align}
where $E_\alpha(x)= \int_1^{\infty} e^{-t x}/t^\alpha dt $ is the exponential integral function \cite{DLMF}.

\begin{figure}
\includegraphics[width=1.0\linewidth]{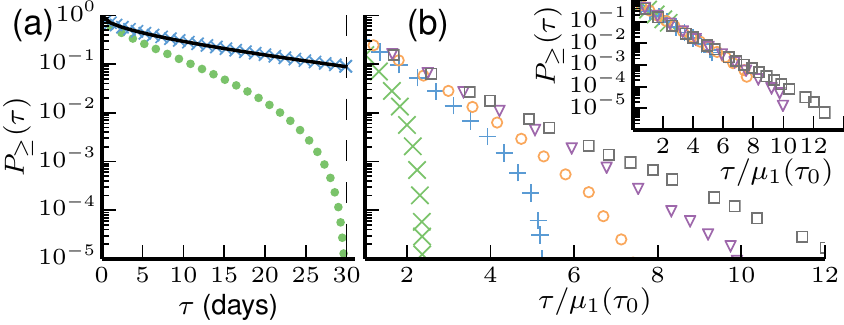}
\includegraphics[width=2.22 in]{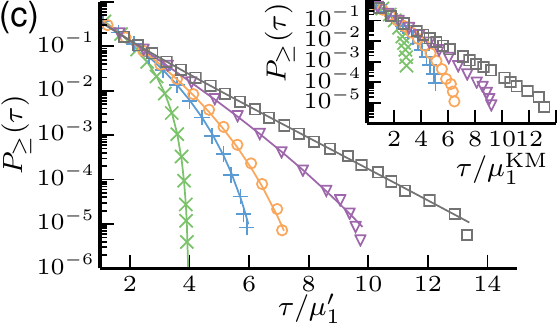}
\caption{Numerical calculations for a model in which we produce the event sequences using the IET distribution $p(\tau | \tau_0) = \frac{1}{\tau_0} f(\tau / \tau_0)$, where $f(\tau)=e^{-\tau}$ and the mean values $\tau_0$ are distributed such that the expected numbers of events satisfy the probability distribution $p(n) \propto n^{-2.5}$ (where $n \geq 1$). 
(a) Cumulative distribution of observed IETs (green dots) and a KM estimate for the cumulative distribution (blue crosses).
The black curve is the theoretical estimate of Eq.~\protect\eqref{eq:iet_collapse_plaw_exp} for the real IET distribution $p(\tau ) \propto E_{\alpha - 2}(\tau / T)$, where $E_n$ is the exponential integral function \protect\cite{DLMF}.
(b) Cumulative distributions of observed IETs when we bin event sequences according to the expected number of observed events $\hat{n}=T/\tau_0$ ({\protect\color{plotcolor1}$\times$}: $\hat{n} \in (2, 3]$, {\protect\color{plotcolor2}$+$}: $\hat{n} \in (5, 6]$, {\protect\color{plotcolor3}$\circ$}: $\hat{n} \in (7, 9]$, {\protect\color{plotcolor4}$\bigtriangledown$}: $\hat{n} \in (13, 25]$, {\protect\color{plotcolor5}$\square$}: $\hat{n} \in (50, 150]$). We skip every other bin to make the figure easier to read, and we divide the IETs in each bin by the mean $\tau_0$ value of the bin $\mu_1(\tau_0)$. In the inset, we show KM estimates for the cumulative distributions IETs of each bin. 
(c) Cumulative distributions of observed IETs when we bin event sequences according to the observed number $n$ of events. We divide the IETs in each bin by the mean observed IET value $\mu_1^\prime$ of the bin.
The lines correspond to IET distributions predicted by Eq.~\protect\eqref{eq:obs_iet_exp_cum} (or to mixtures of them for bins that have event sequences with more than one $n$ value in them).
In the inset, we show KM estimates for the cumulative distributions IETs of each bin and divide the IETs in each bin with $\mu_1^{\rm{KM}}$. 
}
\label{fig:collapse_model}
\end{figure}

In Fig.~\ref{fig:collapse_model}, we show numerical results for the model that we just described. In Fig.~\ref{fig:collapse_model}a, we show both the distribution of observed IETs and a KM estimate that we compute when all of the event sequences are grouped together. It is clear that the observed IETs cannot be used to estimate the real IETs, but the KM estimator performs well in this task.  One can also group event sequences with similar values for the parameter $\tau_0$. Plotting the IET distributions then causes the data to collapse onto a curve that follows the shape given by $f$ if the IET distributions are grouped according to the $\tau_0$ values that were used to generate them and rescaled using the mean of $\tau_0$ values. 
Each group --- and especially the groups with large mean values of $\tau_0$ (i.e., with a small number of events) --- is of course susceptible to finite-size window effects (see Fig~\ref{fig:collapse_model}b), but one can correct for such effects using the same methods as one would use for data produced by a model with a single IET distribution. See the inset of Fig.~\ref{fig:collapse_model}b. 

There is often no way to access the underlying mean IET values $\tau_0$ even if the data is known to be produced by the model that we described above. Instead, one has to estimate $\tau_0$ values from data by calculating the mean IET for each sequence~\cite{Candia2008Uncovering,Goh2008Burstiness,Rybski2012Communication,Karsai2011Small}. This introduces another kind of bias, for which estimators that correct for finite observation windows are not designed.  Our example with exponential $f$ illustrates this situation rather nicely.
In Fig.~\ref{fig:collapse_model}c, we show similar results as in Fig.~\ref{fig:collapse_model}b, except that we group the event sequences using the observed number $n$ of events instead of using $\tau_0$ values of the underlying processes to calculate the expected number of events $\hat{n}=\frac{T}{\tau_0}$. The IET distributions of the event sequences with small numbers of events are not identified correctly as exponential distributions, but instead they follow the distribution defined in Eq.~\eqref{eq:obs_iet_exp} (see below) if one uses the observed number of events to group the event sequences. That is, when grouping event sequences with exactly $n$ events, we find that (1) their IET distributions are independent of the mean rates $\tau_0$ and (2) they cannot be rescaled to follow $f$ even after removing finite-size effects.

\subsection{Deriving observed inter-event time distributions}\label{sec:iet_model}

In this section, we derive a formula for the probability $p^{\prime}(\tau_i, n)$ of observing $\tau_i$ as the $i$th IET in a sequence with exactly $n$ events. We assume that the sequence is produced by a stationary renewal process with an IET distribution of $p(\tau)$ and that we observe it in a finite window that begins at time $0$ and ends at time $T$. We use $p^{\prime}(\tau_i, n)$ to approximate $p^{\prime}(\tau, n)$ when we observe a large number of independent sequences. See Ref.~\cite{Feller1971Introduction} for an introduction to renewal processes.

The probability that the $n$th event after time $0$ takes place at time $t$ is
\begin{align}\label{thisresult}
	p(t,n)=p_R \conv p^{\conv (n-1)}(t)\,,
\end{align}
where $p_r(\tau_r)=\frac{1}{\mu_1} \int_{\tau_r}^{\infty} p(\tau) d\tau$ is the residual waiting-time distribution, $\conv$ is the convolution operator, $x^{\conv y}$ means that $x$ is convolved with itself $y$ times, and $\mu_1$ is the expected IET.
We use Eq.~\eqref{thisresult} to calculate the probability $p^\prime(n)$ of observing exactly $n$ events during a time window of length $T$. The probability $p^\prime(n)$ is equal to the probability that the $n$th event after time $0$ takes place at time $t \leq T$ and the subsequent IET $\tau_{n}$ is larger than $T-t$. 
That is, one can write the probability of observing exactly $n$ events as
\begin{align}
	p^\prime(n) &= \int_0^T p(t,n) \int_{T-t}^\infty p(\tau) d\tau dt \notag \\	
	&= \mu_1 p_R \conv p^{\conv (n-1)}\conv p_R(T)\,.
\label{eq:pn_obs}
\end{align}

We now want to calculate the probability of observing $n$ events when we know the $i$th observed IET $\tau_i$ (where $i \in \{1,\dots,n-1\}$ and $n \geq 2$). We obtain this probability from Eq.~\eqref{eq:pn_obs} by substituting $T$ with $T-\tau_i$ and $n$ with $n-1$ to yield
\begin{align}
	p^\prime(n|\tau_i)=\mu_1 p_R \conv p^{\conv (n-2)}\conv p_R(T-\tau_i)\,.
\end{align}
The joint probability distribution of observing $n$ events with $\tau_i$ as the $i$th IET is thus
\begin{align}
	p^\prime(n,\tau_i)=p^\prime(n|\tau_i)p(\tau_i )\,.
\label{eq:pntau_obs}
\end{align}
Observe that the probability distribution (\ref{eq:pntau_obs}) is independent of the index $i$ as long as $i \in \{1,\dots,n-1\}$.  
By contrast, for a single sequence, the quantities $\tau_i$ and $\tau_j$ (with $i \neq j$) are not independent. However, as long as there are sufficiently many event sequences, we can use Eq.~\eqref{eq:pntau_obs} to approximate the joint distribution of the IETs and the numbers of events.

For a Poisson process, the approximate observed IET distribution given the number of events is
\begin{align}
	p^\prime(\tau | n)=n\frac{(T-\tau)^{n-1}}{T^n}\,.
\label{eq:obs_iet_exp}
\end{align}
The cumulative distribution is thus
\begin{align}
	P^\prime_{\ge }(\tau | n)=\frac{(T-\tau)^{n}}{T^n}\,.
\label{eq:obs_iet_exp_cum}
\end{align}
Note that Eqs.~(\ref{eq:obs_iet_exp}) and (\ref{eq:obs_iet_exp_cum}) are independent of the rate of the Poisson process. We illustrate this independence in Fig.~\ref{fig:collapse_model}c.

\bibliography{ietest}

\begin{thebibliography}{10}
\bibitem{modelfootnote}
Note that we don't use the terminology ``censored on the right'' and ``censored
  on the left'' because we want to avoid confusion with the terms ``right
  censored'' and ``left censored''.
\end{thebibliography}

\end{document}